\documentclass[fleqn,usenatbib]{mnras}

\usepackage[T1]{fontenc}

\usepackage{newtxtext,newtxmath}

\usepackage{graphicx}	
\usepackage{amsmath}	
\usepackage{amssymb}	

\usepackage[modulo,switch]{lineno}

\title[Coupling 1D and 3D simulations on-the-fly, Paper-I]{Coupling 1D stellar evolution with 3D-hydrodynamical simulations on-the-fly I: A~New Standard Solar Model
}

\author[A. C. S. J{\o}rgensen et al.]{
\and
Andreas Christ S{\o}lvsten J{\o}rgensen$^{1}$\thanks{E-mail: acsj@mpa-garching.mpg.de},
Jakob R{\o}rsted Mosumgaard$^{2,1}$\thanks{E-mail: jakob@phys.au.dk},
Achim Weiss,$^{1}$
\and
V\'{i}ctor Silva Aguirre$^{2}$
and
J{\o}rgen Christensen-Dalsgaard$^{2}$
\\
$^{1}$Max-Planck-Institut f\"ur Astrophysik, Karl-Schwarzschild-Str. 1, D-85748 Garching, Germany \\
$^{2}$Stellar Astrophysics Centre (SAC), Department of Physics and Astronomy, Aarhus University, Ny Munkegade 120, DK-8000 Aarhus C, Denmark\\
}
\date{Accepted XXX. Received YYY; in original form ZZZ}

\pubyear{2018}

\begin{document}
\label{firstpage}
\pagerange{\pageref{firstpage}--\pageref{lastpage}}
\maketitle

\begin{abstract}
Standard 1D stellar evolution models do not correctly reproduce the structure of the outermost layers of stars with convective envelopes. This has been a long-standing problem in stellar modelling affecting both the predicted evolutionary paths and the attributed oscillation frequencies, and indirectly biasing numerous quantities derived from stellar evolution calculations. We present a novel method that mostly eliminates these structural defects by appending mean 3D simulations of stellar envelopes. In contrast to previous attempts we impose the complete structure derived from 3D simulations at each time step during the entire evolution. For this purpose, we interpolate in grids of pre-computed 3D simulations and use the resulting structure as boundary conditions, in order to solve the stellar structure equations for the 1D interior at each time step. Our method provides a continuous transition in many quantities from the interior to the imposed interpolated 3D surface layers. We present a solar calibration model and show that the obtained structure of the surface layers reliably mimics that of the underlying 3D simulations for the present Sun. Moreover, we perform a helioseismic analysis, showing that our method mostly eliminates the structural contribution to the discrepancy between model frequencies and observed p-mode frequencies.
\end{abstract}





\section{Introduction}
\label{sec:intro}

Stellar evolution codes do not yield the correct outermost structure of stars with convective envelopes. Instead, 1D stellar models employ {\color{black}1D model atmospheres, such as plane-parallel Eddington grey atmospheres, and parametrizations of superadiabatic convection, such the mixing length theory by \cite{Bohm-Vitense1958} (MLT). This treatment of convection necessitates the calibration of a free parameter, the mixing length ($\alpha_\textsc{mlt}$), which is usually done based on the present Sun. This is a well-known weakness of MLT, since the extent to which the calibrated value is applicable for other evolutionary stages is disputed.} Furthermore, the inadequate modelling of the surface layers results in a systematic offset between observations and the predicted p-mode frequencies, the so-called surface effect \citep{Brown1984,Christensen-Dalsgaard1988}. When interpreting asteroseismic data from, e.g., the CoRoT \citep{Baglin2009} and the \textit{Kepler} \citep{Borucki2010x} space missions, most  authors account for this discrepancy by using empirical corrections based on the Sun \citep{Kjeldsen2008,Ball2014}. 

\cite{Schlattl1997} and \cite{Rosenthal1999} were among the first to include data from 2D and 3D hydrodynamic simulations of stellar envelopes, in order to improve the structure of the surface layers. Recently, several authors have addressed the problem similarly by substituting the outermost envelope with mean structures from 3D hydrodynamic simulations \citep{Piau2014, Sonoi2015, Ball2016, Magic2016,Joergensen2017, Trampedach2017}. In all mentioned cases, this substitution has been performed for a given final structure, i.e. the 3D simulations have not been included throughout the evolution. This substitution is known as patching. As elaborated upon by \cite{Joergensen2017} --- hereafter J17 ---, the fact that patching is performed after a traditional 1D evolutionary calculation {\color{black} has been performed} can lead to inconsistencies and discontinuities in the patched models. Nevertheless, as shown by \cite{Houdek2017}, this approach does mend the structural inadequacies of stellar models and results in model frequencies that are in good agreement with observations once so-called modal effects including non-adiabatic energetics have been taken into account.

In order to include information from 3D simulations during the entire evolution, \cite{Trampedach2014a,Trampedach2014b} have computed $T(\tau$) relations and calibrated mixing lengths for a grid of 3D models. Here $T(\tau)$ denotes the temperature as a function of optical depth. While this is a step forward, an implementation by \cite{Mosumgaard2018a} shows that this parametrization does not account for the structural surface effect. The reason for this inability to mimic the correct structure is that the calibrated mixing length only reproduces the mean temperature gradient of the envelope and not the actual temperature gradient as a function of depth. {\color{black}Furthermore, these 1D models neither account properly for turbulent pressure nor for so-called convective back-warming \cite[e.g.][]{Trampedach2017}.}

In this paper, we present the solar calibration model for which we have substituted the outermost layers with a mean 3D structure, throughout the entire evolution of the star, adjusting the interior model accordingly in each iteration at every time step. This solar model has been obtained using the usual solar calibration procedure, i.e. a first-order Newton iteration scheme \citep[cf.][]{Weiss2008}. We compare the obtained structure with the employed 3D simulations and present p-mode frequencies.

\section{Implementation}
\label{sec:implement}

We compute stellar structure models, employing the Garching Stellar Evolution Code \citep[\textsc{garstec,}][]{Weiss2008}. In each iteration, at every time step, the pressure and temperature stratifications of the outermost layers have been adopted from interpolated mean 3D hydrodynamic simulations of stellar atmospheres --- denoted $\langle 3\mathrm{D} \rangle$-envelopes hereafter. For this purpose, we have used the Stagger-grid \citep{Magic2013} and the interpolation method presented by J17. We have only included models with solar metallicity: For the sake of consistency between the interior model and the Stagger-grid, we use the composition published by \cite{Asplund2009} (AGSS09). Of the corresponding 29 3D simulations in the Stagger-grid we have excluded one that is not fully relaxed. 

\textsc{garstec} determines stellar structure models from the centre to the outer boundary by solving the stellar structure equations, i.e. a set of coupled differential equations, using the Henyey scheme. Usually, the outer boundary is placed at the photosphere, and the boundary conditions are obtained from Stefan-Boltzmann's law and from an integration of an Eddington grey atmosphere.

\begin{figure}
  \centering
  \includegraphics[width=\linewidth]{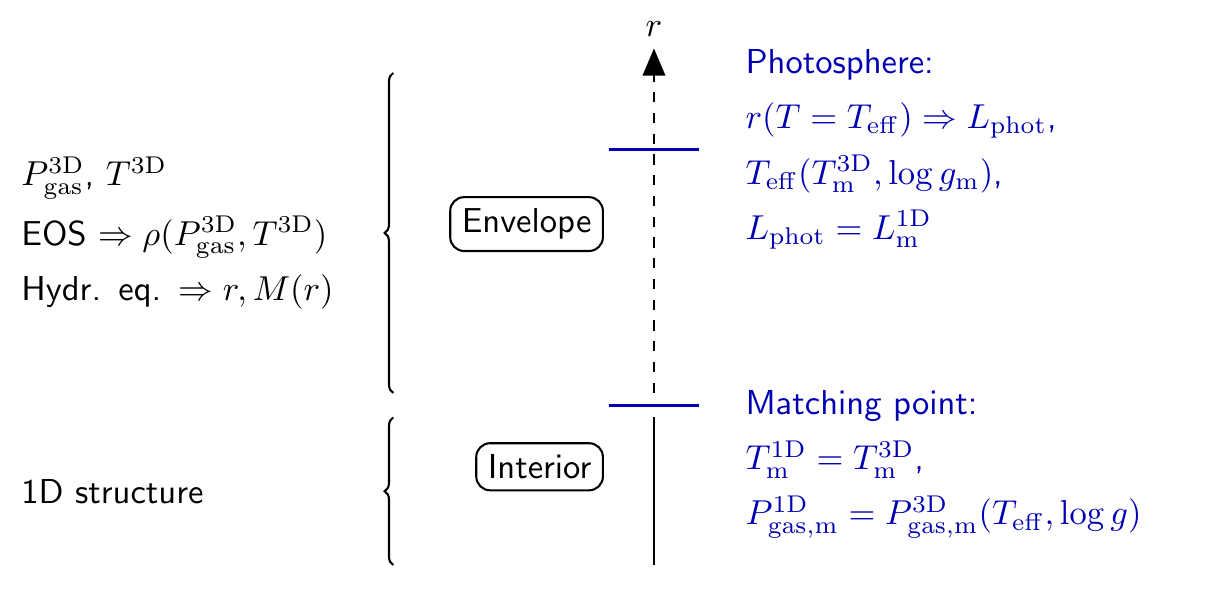}
  \caption{Schematic overview of the implementation of averaged 3D-envelopes in \textsc{garstec}. The interior structure is found, using the Henyey scheme, based on the stellar structure equations. The appended envelope is adopted from interpolated 3D simulations and yields the outer boundary conditions of the interior. The subscript 'm' refers to values at the matching point. See text for details.
  }
  \label{fig:illustr}
\end{figure}

In our implementation we supply the two required outer boundary conditions by employing interpolated $\langle \mathrm{3D} \rangle$-envelopes. The method is sketched in Fig.~\ref{fig:illustr}. As illustrated in this figure, we supply the boundary conditions of the interior model far below the photosphere, at the so-called matching point. Outside of the matching point, the temperature ($T$) as a function of gas pressure ($P_\mathrm{gas}$) is taken directly from interpolated $\langle \mathrm{3D} \rangle$-structures. We will use {\color{black}t}he superscripts '1D' and '3D' to indicate whether the value refers to the interior model or the $\langle \mathrm{3D} \rangle$-envelope, respectively.

As suggested by J17, we interpolate $P^{\mathrm{3D}}_\mathrm{gas}$ and $T^{\mathrm{3D}}$ in the $(T_\mathrm{eff},\log g)$-plane, scaling both quantities before the interpolation. Here $T_\mathrm{eff}$ and $\log g$ denote the effective temperature and logarithm of the gravitational acceleration in base ten, respectively. The scaling factor is set to be the corresponding value ($P^\mathrm{3D}_\mathrm{gas,jump}$) at the location of the minimum in $\mathrm{d}\ln{\rho^\mathrm{3D}}/\mathrm{d}\ln{P^\mathrm{3D}_\mathrm{gas}}$, i.e. at the density jump near the surface, where the density inversion takes place in later evolutionary stages. This scaling ensures that abrupt changes in all relevant quantities near the surface take place at the same scaled pressure, making the interpolation more reliable: the scaled stratifications look rather similar across the $(T_\mathrm{eff},\log g)$-plane (cf. Fig.~2 in J17) --- that is, some aspects of the Stagger-grid envelopes are homologous. Moreover, the scaling factors behave rather well as a function of $T_\mathrm{eff}$ and $\log g$ ($P^\mathrm{3D}_\mathrm{gas,jump}$ behaves linearly, cf. Fig.~3 in J17). After the interpolation, the scaling is inverted.

During an initial phase on the pre-main sequence, the matching point is moved inwards, starting near the photosphere. This is done for convergence purposes and implies that the scaled gas pressure, i.e. $P^{\mathrm{3D}}_\mathrm{gas}/P^{\mathrm{3D}}_\mathrm{gas,jump}$ at the matching point increases. For the remaining evolution, the scaled gas pressure at the matching point is kept fixed. For a given value of the scaled gas pressure, any combination of $T_\mathrm{eff}$ and $\log g$ corresponds to a unique value of $T^{\mathrm{3D}}$ at the matching point: $T^\mathrm{3D}_\mathrm{m}$. Here, the subscript 'm' refers to the value at the matching point (cf. Fig.~\ref{fig:illustr}). 

In order to compute the gas pressure and the temperature of the $\langle \mathrm{3D} \rangle$-envelope at the matching point, we need to infer $T_\mathrm{eff}$. We do this by enforcing a match in $T$ at the matching point, i.e. by inverting the problem: we evaluate the effective temperature that corresponds to $T^\mathrm{3D}_\mathrm{m}=T^\mathrm{1D}_\mathrm{m}$ and $\log g_\mathrm{m}$, by interpolation in the Stagger-grid.

Having established $T_\mathrm{eff}$, we then compute the appropriate scaling factor for the gas pressure by interpolation. This gives us the pressure at the matching point ($P^\mathrm{3D}_\mathrm{gas,m}$) of the interpolated 3D structure, which is to be compared with the gas pressure predicted by the interior model at the matching point ($P^\mathrm{1D}_\mathrm{gas,m}$). This yields the first of our two boundary condition: $P^\mathrm{1D}_\mathrm{gas,m}$ must match $P^\mathrm{3D}_\mathrm{gas,m}$. The Henyey scheme will iteratively adjust the interior structure to achieve this, which will lead to changes in $P^\mathrm{1D}_\mathrm{gas,m}$. For each Henyey iteration, $T_\mathrm{eff}$ and hence $P^\mathrm{3D}_\mathrm{gas,m}$ are also re-evaluated until a match is found.

Above the matching point $P_\mathrm{gas}$ is determined based on the computed scaling factor, and $T$ as a function $P_\mathrm{gas}$ is derived by interpolation in the $\langle 3\mathrm{D \rangle}$-envelope: thus, we append $P_\mathrm{gas}$ and $T$ from an interpolated $\langle 3\mathrm{D} \rangle$-envelope in each iteration. The density ($\rho$) at each mesh point in the appended envelope is then computed by the stellar evolution code from the equation of state (EOS). We note that the EOS used in the stellar evolution code to obtain $\rho$ is not identical to the EOS used in the 3D simulations. Nevertheless, as discussed in the next section, $\rho$ is reproduced with high accuracy. The radius ($r$) is computed iteratively from hydrostatic equilibrium:
\begin{equation}
\frac{\mathrm{d}P}{\mathrm{d}\mathrm{r}} = - \frac{GM(r)}{r^2} \rho. \label{eq:hydrostatic}
\end{equation}
Here $G$ is the gravitational constant, and $M(r)$ denotes the mass within $r$, which can be derived iteratively from $\rho$ and $r$. $P$ denotes the total pressure including the turbulent pressure. In this paper, we ignore the contribution from the turbulent pressure. The remaining thermodynamic quantities in the envelope, including the first adiabatic index ($\Gamma_1$), are determined from the EOS. Since the surface layers are convective, the composition of the envelope is set to be the same as in the outermost point of the interior model. In this way, we end up with a pseudo-$\langle 3\mathrm{D} \rangle$-structure beyond the matching point. 

Having established $r$ for each mesh point, we determine $r(T=T_\mathrm{eff})$ by interpolation, in order to compute the associated luminosity ($L_\mathrm{phot}$) from Stefan-Boltzmann's law, as this law holds true at the photosphere. Under the assumption that the energy generation in the appended envelope is negligible, we compare the result of Stefan-Boltzmann's law ($L_\mathrm{phot}$) with the luminosity at the matching point ($L^\mathrm{1D}_\mathrm{m}$). This is the second boundary condition for the Henyey scheme: $L_\mathrm{phot}$ must match $L^\mathrm{1D}_\mathrm{m}$. 

When an equilibrium structure is reached, the chemical composition is evolved in time, and a new equilibrium structure is evaluated by the Henyey scheme. 

\section{A solar calibration}
\label{sec:solarcal}

We have performed a solar model calibration, using the Stagger-grid as the boundary condition and appending a $\langle \mathrm{3D} \rangle$-envelope as described above. We have used the OPAL opacities \citep{Iglesias1996} in combination with the low-temperature opacities by \cite{Ferguson2005}. We use AGSS09 and the FreeEOS by A.~W. Irwin\footnote{The code can be found on http://freeeos.sourceforge.net/}, which is briefly described by \cite{Cassisi2003}. Our solar model includes diffusion of H, He, C, N, O, Ne, Mg, Si and Fe. As we do not interpolate in the composition of the 3D simulations, only the envelope of the final model of the present Sun has the same composition as the underlying 3D simulation.

After an initial phase on the pre-main sequence the logarithm of the scaled pressure, i.e. $\log_{10}(P^\mathrm{3D}_\mathrm{gas,m}/P^\mathrm{3D}_\mathrm{gas,jump})$, is fixed at $1.20$ throughout the entire subsequent evolution. For the associated model of the current Sun, the chosen scaled pressure corresponds to a temperature of $1.41\times 10^{4}\,\mathrm{K}$ and a depth of $0.94\, \mathrm{Mm}$ below the photosphere. The matching point should be placed as deep within the 1D model as possible, since the stratification becomes more adiabatic with increasing depth. The chosen scaled pressure at the matching point is a compromise between this requisite and the limitations of the interpolation scheme imposed by the resolution of the Stagger-grid and by the depths of Stagger-grid simulations (cf.~J17).

The final calibrated solar model matches the current solar radius ($R_\odot$) and luminosity ($L_\odot$) as well as the desired abundance of heavy elements relative to hydrogen at the solar surface ($Z_\mathrm{S}/X_\mathrm{S}=0.01828$) to better than a relative accuracy of $10^{-5}$. The solar values have been chosen in such a way as to recover the effective temperature of the solar Stagger-grid envelope: $T_\mathrm{eff}=5768.5\,$K.

Our implementation of $\langle 3\mathrm{D} \rangle$-envelopes in stellar models yields a mixing length ($\alpha_\textsc{mlt}$) that is roughly a factor of two higher than the value obtained, when using an Eddington grey atmosphere. This reflects the fact that the mixing length is calibrated to reproduce $R_\odot$: When using the Stagger-grid as boundary conditions, the mixing length directly affects only the extent of a superadiabatic layer that is much thinner than the corresponding layer in standard solar models, since the extent of the region beyond the matching point is dictated by the interpolated $\langle 3\mathrm{D} \rangle$-envelope. The change in $\alpha_\textsc{mlt}$ hence reflects the fact that 3D atmospheres are more extended than their 1D counterparts{\color{black}, due to turbulent pressure and convective back-warming}. A similar result was obtained by \cite{Schlattl1997}, who pointed out that a spatially variable mixing length must be introduced, when appending more realistic outer layers, in order to reproduce the $T$ as a function $P_\mathrm{gas}$. {\color{black}Physically, $\alpha_\textsc{mlt}$ reproduces the mean temperature gradient between the top and bottom of the superadiabatic layer. With increasing matching depth, the average temperature gradient of the region modelled by MLT will be closer to being adiabatic, and hence $\alpha_\textsc{mlt}$ should increase accordingly.}

\begin{table}
  \caption{Derived initial parameters as well as the helium surface abundance and the radius of the base of the convective envelope for two solar calibrations, using an Eddingron atmosphere and the Stagger-grid as boundary conditions, respectively.}
  \label{tab:suncal}
  \begin{tabular}{lllllll}
    \hline
    Bound. con. & $\alpha_\textsc{mlt}$ & $Y_\mathrm{i}$ & $Z_\mathrm{i}$ & $Y_\mathrm{S}$ & $\frac{r_\mathrm{cz}}{R_\mathrm{\odot}}$ \\
    \hline
    Eddington & 1.78 & 0.2653 & 0.0152 & 0.2343 & 0.724{\color{black}4} \\[2pt]
    $\langle 3\mathrm{D} \rangle$-envelope & 3.30 & 0.2652 & 0.0152 & 0.2343 & 0.724{\color{black}3} \\[2pt]
    \hline
  \end{tabular}
\end{table}

As will be discussed in a companion paper (Paper-II, Mosumgaard et al., in prep.) the obtained value of $\alpha_\textsc{mlt}$ is affected by the matching depth. However, in Paper-II, this is shown not to shift the evolutionary track significantly if the matching point is placed sufficiently deep within the superadiabatic region.

The obtained $\alpha_\textsc{mlt}$ for the presented solar model is listed in Table~\ref{tab:suncal} alongside other quantities that result from the solar calibration. $Y_\mathrm{S}$ in Table~\ref{tab:suncal} denotes the surface helium abundance and should be compared to the seismic value obtained by \cite{Basu2004}: $0.2485\pm 0.0035$. $r_\mathrm{cz}$ denotes the radius of the base of the convection zone and should be compared to the value that has been seismically inferred by \cite{Basu1997}: $0.713\pm 0.001 R_\odot$. For both quantities, the discrepancy between the model and observations can be attributed to the use of AGSS09 \citep{Serenelli2009}.

Table~\ref{tab:suncal} also contains the corresponding values for a standard solar calibration that employs an Eddington grey atmosphere. As can be seen from Table~\ref{tab:suncal}, our method, involving  $\langle 3\mathrm{D} \rangle$-envelopes, does not affect $Y_\mathrm{S}$ or $r_\mathrm{cz}$ significantly. {\color{black}We attribute the fact that both solar calibration models lead to nearly identical radii at the base of the convection zone to their very similar boundary conditions.}

\subsection{Structure of outermost layers}
\label{sec:solarcal_struc}

\begin{figure}
  \centering
  \includegraphics[width=\linewidth]{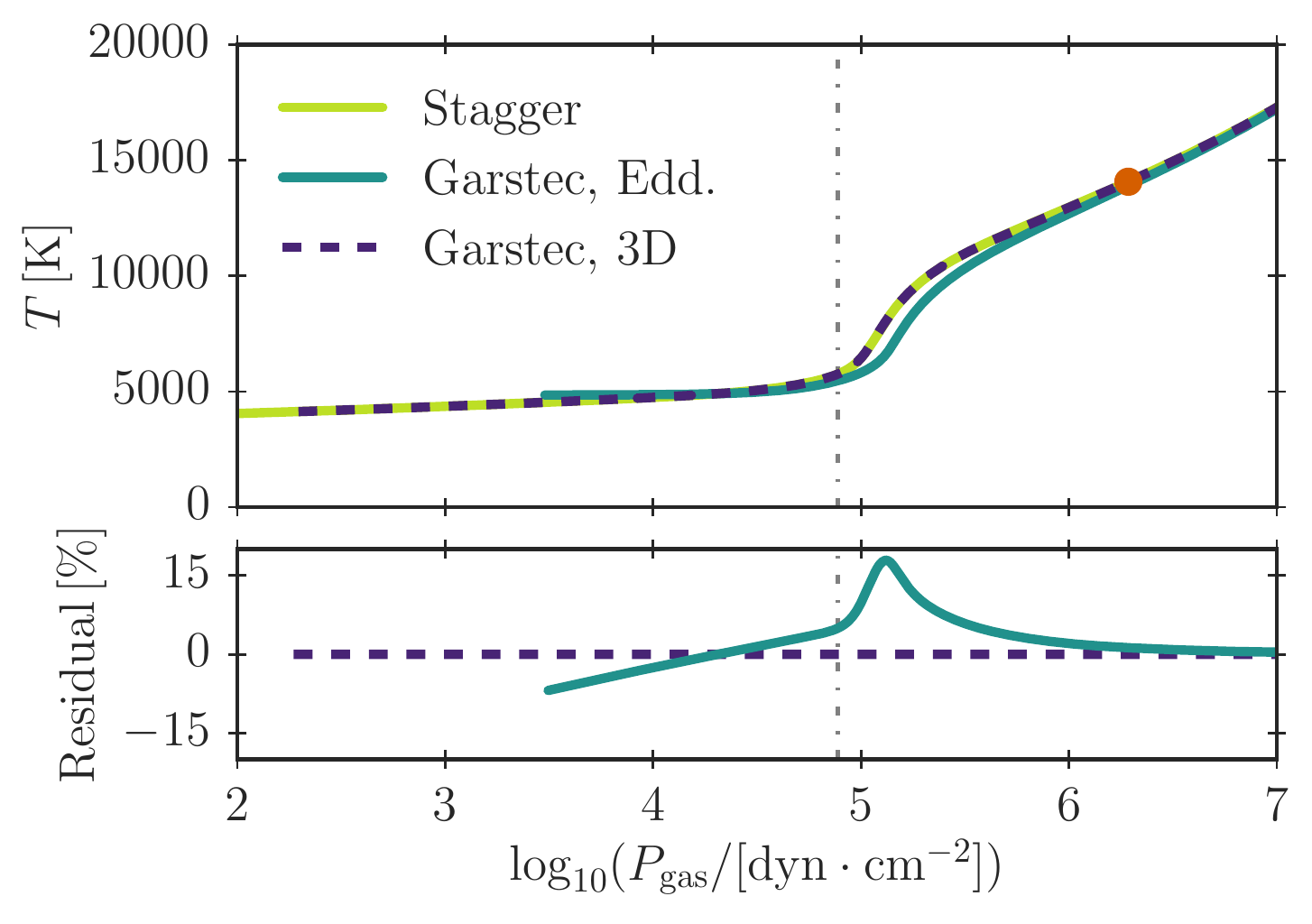}
  \caption{\textbf{Upper panel:} Temperature as function of gas pressure for the $\langle 3\mathrm{D} \rangle$ solar envelope in the Stagger-grid as well as the two solar calibrations listed in Table~\ref{tab:suncal}. \textbf{Lower panel:} Residuals between the solar Stagger-grid envelope and each solar calibration relative to the corresponding value in the Stagger-grid envelope. The vertical grey line indicates the position of the photosphere. The read dot denotes the location of the matching point.
  }
  \label{fig:tvsP}
\end{figure}

\begin{figure}
  \centering
  \includegraphics[width=\linewidth]{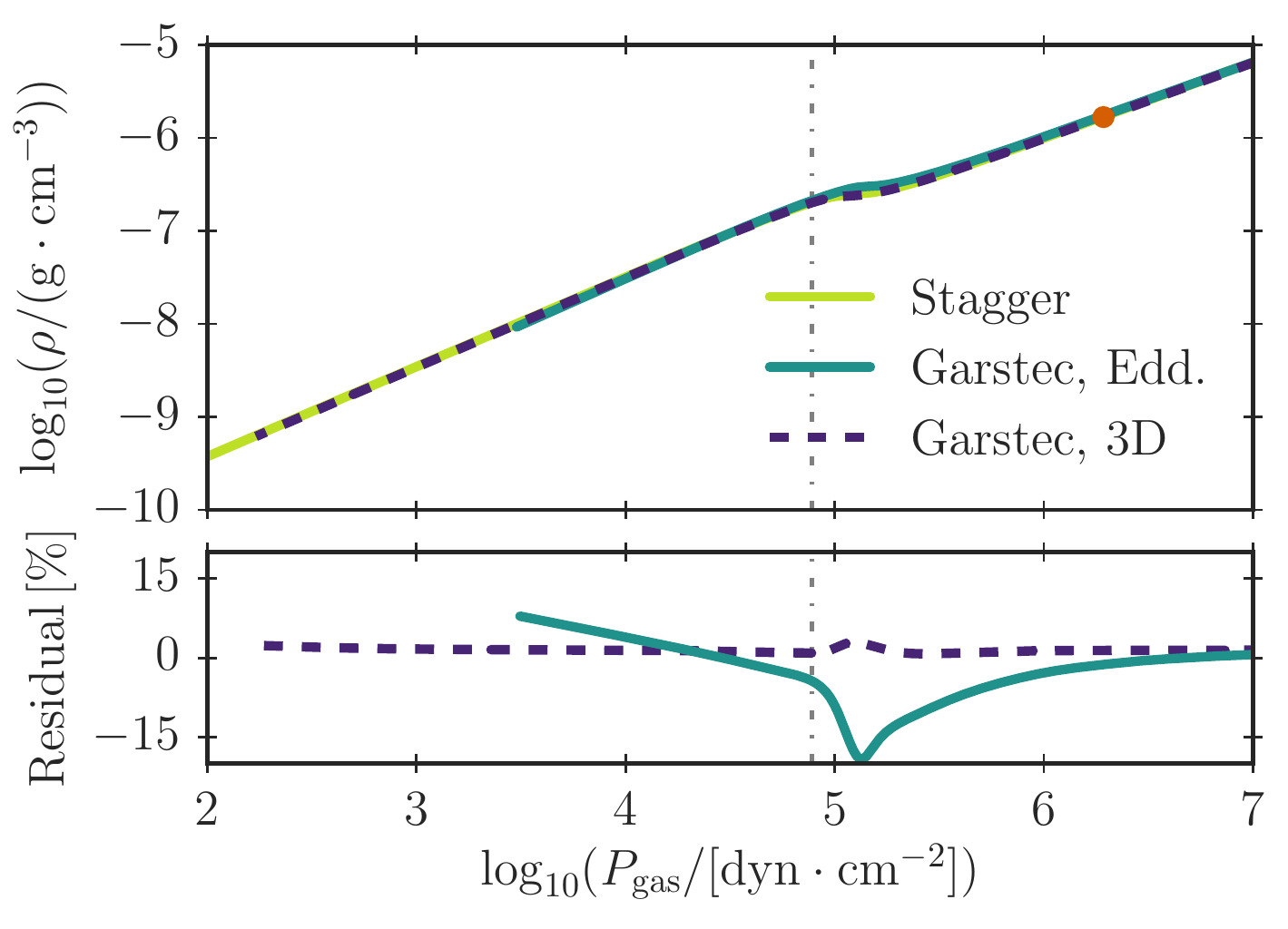}
  \caption{As Fig.~\ref{fig:tvsP} but for the density as a function of gas pressure.
  }
  \label{fig:rhovsP}
\end{figure}

\begin{figure}
  \centering
  \includegraphics[width=\linewidth]{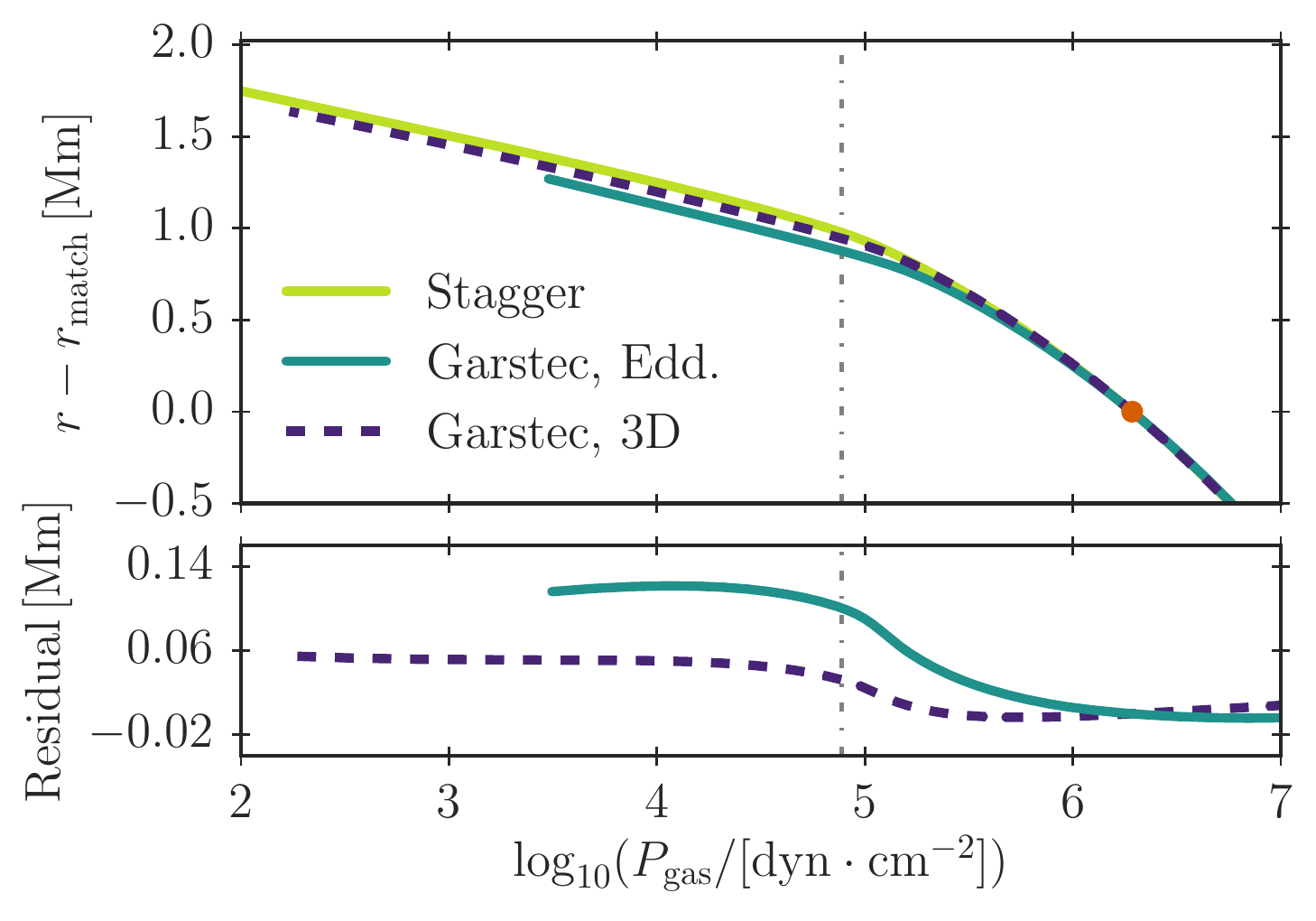}
  \caption{As Fig.~\ref{fig:tvsP} but for the height above the matching point as a function of gas pressure. Here we show the absolute residuals.
  }
  \label{fig:rvsP}
\end{figure}

We now compare the structure obtained by \textsc{garstec} with that of the 3D simulation in the Stagger-grid, whose global parameters correspond to the solar values. The results are shown in Figs~\ref{fig:tvsP}-\ref{fig:rvsP}. The figures include both \textsc{garstec} models from Table~\ref{tab:suncal}: one uses the Stagger-grid as its boundary conditions and appends the corresponding $\langle \mathrm{3D} \rangle$-envelope at each time step, while the other employs an Eddington grey atmosphere.

{\color{black}Fig.~\ref{fig:tvsP} shows $T(P_\mathrm{gas})$. By construction our new scheme reproduces this stratification correctly.} From Fig.~\ref{fig:rhovsP}, we see that the EOS used by the 1D evolution code closely recovers $\rho(P_\mathrm{gas})$, although the density is systematically $1-2$\,\% too {\color{black}low}. 
This may partly reflect differences in the EOS used by the 1D and 3D codes. Furthermore, since $\rho$ is a non-linear function of $P_\mathrm{gas}$ and $T$, the geometrical mean of the density in the Stagger-grid is not expected to correspond to the density derived from the average of $P_\mathrm{gas}$ and $T$ (cf. R. Collet, private communication).

{\color{black}Fig.~\ref{fig:rvsP} shows that our method closely\footnote{We note that $\log g$ is assumed to be constant in the Stagger-grid envelope, in contrast to the 1D simulation. This slightly skews the comparison in $r$.} reproduces the depth of the underlying 3D envelope as a function of gas pressure, despite the neglect of turbulent pressure and the discrepancy in $\rho$}. Thus, our new method reproduces the expected $\langle 3\mathrm{D} \rangle$-structure from the Stagger-grid very well, without the need for post-evolutionary patching. The Eddington grey atmosphere shows much larger residuals, as was to be expected, since we modify the outer convective layers only.

\subsection{Oscillation frequencies}
\label{sec:solarcal_osc}

\begin{figure}
  \centering
  \includegraphics[width=\linewidth]{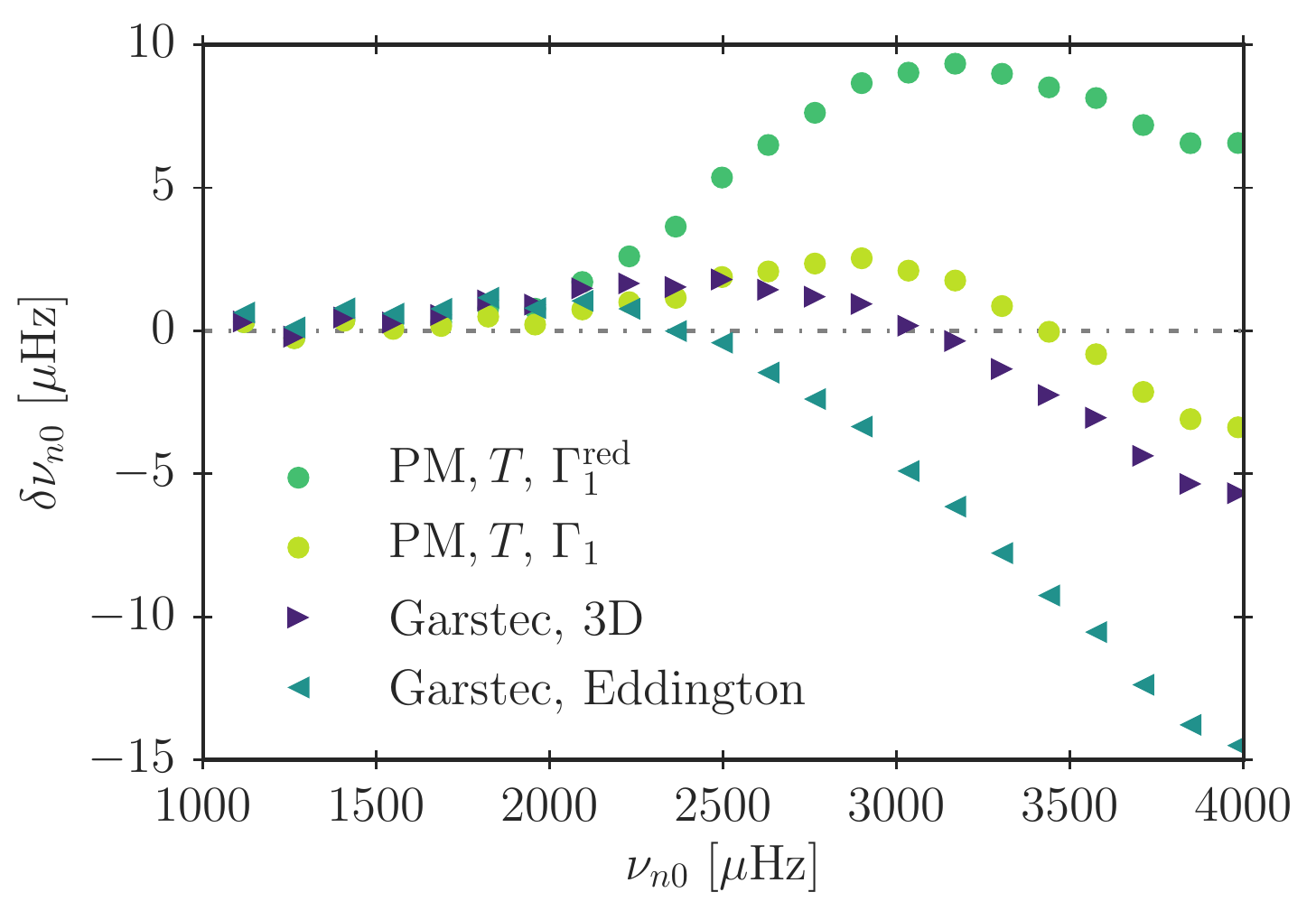}
  \caption{Frequency difference between BiSON observations and the solar models listed in Table~\ref{tab:suncal}. We only include radial modes ($\ell=0$) with frequencies above $1000\,\mu\mathrm{Hz}$.  
  }
  \label{fig:Freqnew}
\end{figure}

Having obtained a structure that closely resembles the mean stratification of 3D simulations, we have computed stellar oscillation frequencies using the Aarhus adiabatic oscillation package \citep[\textsc{adipls}][]{Christensen-Dalsgaard2008a}. Fig.~\ref{fig:Freqnew} shows the frequency differences, $\delta \nu_{n\ell}$, between the predicted model frequencies and observations from the Birmingham Solar Oscillation Network \citep[BiSON][]{Broomhall2009,Davies2014}. Here $\nu_{n\ell}$ denotes the frequencies, $n$ is the radial order and $\ell$ is the degree.

For comparison, we have also included a patched model (PM, cf. J17) in Fig.~\ref{fig:Freqnew}, i.e. a model, for which we have substituted the outermost layers {\color{black}\textit{of the present Sun}} with the {\color{black}entire} mean structure of the solar Stagger-grid envelope. We have used $T$ to determine the radius at which to patch (cf. J17). The patch has been performed at a depth of $2.1\,\mathrm{Mm}$ below the surface, i.e. further down within the nearly adiabatic region than the matching point. The patched model has been constructed such that it has the same interior as the {\color{black}final} solar calibration model, whose outer layers are dictated by the Stagger-grid. {\color{black}As opposed to the models constructed on the fly, the patched model hence includes turbulent pressure in the envelope --- that is, turbulent pressure was neglected during the entire evolution and is not taken into account below the patched envelope.}

In order to compute frequencies for the patched model, one must account for the turbulent pressure, which has been included in the patched envelope, in the oscillation equations. As elaborated upon by \cite{Houdek2017}, this can be achieved by adjusting $\Gamma_1=(\partial \ln P_\mathrm{gas}/\partial \ln \rho)_\mathrm{ad}$ by a factor of $P_\mathrm{gas}/P$: $\Gamma_1^\mathrm{red} = (P_\mathrm{gas}/P)\Gamma_1$. The resulting quantity is referred to as the reduced adiabatic index, $\Gamma_1^{\mathrm{red}}$ \citep{Rosenthal1999}. The associated frequencies are included in Fig~\ref{fig:Freqnew}.

However, some authors \citep{Ball2016, Magic2016, Joergensen2017} do not use $\Gamma_1^{\mathrm{red}}$ but compute frequencies of patched models based on $\Gamma_1$. This amounts to the assumption that the turbulent pressure reacts in the same way as the gas pressure to density perturbations.
To facilitate an easy comparison with these authors, we have recomputed the frequencies employing $\Gamma_1$ and included these in Fig~\ref{fig:Freqnew}.

As can be seen from the figure, our new method leads to model frequencies that are very similar to the frequencies obtained from patched models when using $\Gamma_1$ in the frequency calculations: at $4000\,\mu\mathrm{Hz}$ the remaining frequency difference between our solar calibration and observations is only $6\,\mu\mathrm{Hz}$, while the frequency difference is $15\,\mu\mathrm{Hz}$ at the same frequency, when using an Eddington grey atmosphere. With our new method, we are thus able to significantly reduce the structural contribution to the surface effect. 

The remaining frequency differences between the patched models and our solar calibration model that employs the Stagger-grid as its boundary conditions can partly be attributed to the neglect of turbulent pressure and the {\color{black}discrepancies in the density stratification.}. 
Furthermore, this frequency differences will, at least to some extent, reflect the patching and matching depths. As will be discussed in Paper-II, however, the frequencies are rather insensitive to the scaled pressure of the matching point if the matching point is placed sufficiently deep within the superadiabatic region.

\section{Conclusion}
\label{sec:conclusion}

We have presented a novel method for including the mean pressure and temperature stratifications from 3D simulations directly in 1D stellar evolution codes and exploit this to adjust the structure at every time step. This has been achieved without the need of parametrizations or post-evolutionary patching. It is the first computation of a solar model, for which the structure from 3D simulations has been fully accounted for on the fly. The structure of the resulting 1D models is in very good agreement with the underlying 3D simulations, despite the neglect of turbulent pressure, and despite the fact that only a limited amount of information is taken from the 3D models.

We find that our new method is largely able to eliminate the structural surface effect, leading to very promising p-mode frequencies. Modal effects have not been accounted for. A direct comparison with post-evolutionarily patched models show only small frequency differences. We largely attribute the remaining differences to the neglect of turbulent pressure.

In the case of the present Sun, our analysis confirms that post-evolutionary patching provides an adequate structural correction. A disadvantage of patching, however, is that it relies on the assumption that the evolutionary tracks are unaltered by the use of crude outer boundary conditions and mixing length theory. Our method, on the other hand, overcomes this deficiency by adjusting the structure of stellar models on the fly based on 3D simulations. We address the impact of our new method on the evolutionary tracks in detail in Paper-II. In the companion paper, we will go beyond the present Sun. This includes the first asteroseismic analysis of stars in the \textit{Kepler} field and an investigation of how well our method reproduces the correct structures of 3D-envelopes throughout $(T_\mathrm{eff},\log g)$-plane. In future work, we also plan to alter the presented method: we intend to include turbulent pressure and take composition changes into account by interpolation in $\mathrm{[Fe/H]}$.

\section*{Acknowledgements}

We thank R. Collet, Z. Magic, H. Schlattl and R. Trampedach for their collaboration as well as the Max Planck Institut f\"{u}r Astrophysik, the Danish National Research Foundation (Grant DNRF106) and Villum Fonden (research grant 10118) for funding. 




\bibliographystyle{mnras}
\bibliography{manual_refs,mendeley_export}

\begin{thebibliography}{}
\makeatletter
\relax
\def\mn@urlcharsother{\let\do\@makeother \do\$\do\&\do\#\do\^\do\_\do\%\do\~}
\def\mn@doi{\begingroup\mn@urlcharsother \@ifnextchar [ {\mn@doi@}
  {\mn@doi@[]}}
\def\mn@doi@[#1]#2{\def\@tempa{#1}\ifx\@tempa\@empty \href
  {http://dx.doi.org/#2} {doi:#2}\else \href {http://dx.doi.org/#2} {#1}\fi
  \endgroup}
\def\mn@eprint#1#2{\mn@eprint@#1:#2::\@nil}
\def\mn@eprint@arXiv#1{\href {http://arxiv.org/abs/#1} {{\tt arXiv:#1}}}
\def\mn@eprint@dblp#1{\href {http://dblp.uni-trier.de/rec/bibtex/#1.xml}
  {dblp:#1}}
\def\mn@eprint@#1:#2:#3:#4\@nil{\def\@tempa {#1}\def\@tempb {#2}\def\@tempc
  {#3}\ifx \@tempc \@empty \let \@tempc \@tempb \let \@tempb \@tempa \fi \ifx
  \@tempb \@empty \def\@tempb {arXiv}\fi \@ifundefined
  {mn@eprint@\@tempb}{\@tempb:\@tempc}{\expandafter \expandafter \csname
  mn@eprint@\@tempb\endcsname \expandafter{\@tempc}}}

\bibitem[\protect\citeauthoryear{Asplund, Grevesse, Sauval  \& Scott}{Asplund
  et~al.}{2009}]{Asplund2009}
Asplund M.,  Grevesse N.,  Sauval A.~J.,   Scott P.,  2009, \mn@doi [ARA\&A]
  {10.1146/annurev.astro.46.060407.145222}, 47, 481

\bibitem[\protect\citeauthoryear{{Baglin}, {Auvergne}, {Barge}, {Deleuil},
  {Michel}  \& {CoRoT Exoplanet Science Team}}{{Baglin}
  et~al.}{2009}]{Baglin2009}
{Baglin} A.,  {Auvergne} M.,  {Barge} P.,  {Deleuil} M.,  {Michel} E.,   {CoRoT
  Exoplanet Science Team} 2009, \mn@doi [Proceedings of the International
  Astronomical Union] {10.1017/S1743921308026252}, 253, 71

\bibitem[\protect\citeauthoryear{Ball \& Gizon}{Ball \& Gizon}{2014}]{Ball2014}
Ball W.~H.,  Gizon L.,  2014, \mn@doi [A\&A] {10.1051/0004-6361/201424325},
  568, A123

\bibitem[\protect\citeauthoryear{Ball, Beeck, Cameron  \& Gizon}{Ball
  et~al.}{2016}]{Ball2016}
Ball W.~H.,  Beeck B.,  Cameron R.~H.,   Gizon L.,  2016, \mn@doi [A\&A]
  {10.1051/0004-6361/201628300}, 592, A159

\bibitem[\protect\citeauthoryear{Basu \& Antia}{Basu \& Antia}{1997}]{Basu1997}
Basu S.,  Antia H.~M.,  1997, \mn@doi [MNRAS] {10.1093/mnras/287.1.189}, 287,
  189

\bibitem[\protect\citeauthoryear{Basu \& Antia}{Basu \& Antia}{2004}]{Basu2004}
Basu S.,  Antia H.~M.,  2004, \mn@doi [ApJ] {10.1086/421110}, 606, L85

\bibitem[\protect\citeauthoryear{B{\"{o}}hm-Vitense}{B{\"{o}}hm-Vitense}{1958}]{Bohm-Vitense1958}
B{\"{o}}hm-Vitense E.,  1958, Zeitschrift f{\"{u}}r Astrophysik, 46

\bibitem[\protect\citeauthoryear{{Borucki} et~al.,}{{Borucki}
  et~al.}{2010}]{Borucki2010x}
{Borucki} W.~J.,  et~al., 2010, \mn@doi [Science] {10.1126/science.1185402},
  \href {http://adsabs.harvard.edu/abs/2010Sci...327..977B} {327, 977}

\bibitem[\protect\citeauthoryear{Broomhall, Chaplin, Davies, Elsworth,
  Fletcher, Hale, Miller  \& New}{Broomhall et~al.}{2009}]{Broomhall2009}
Broomhall A.-M.,  Chaplin W.~J.,  Davies G.~R.,  Elsworth Y.,  Fletcher S.~T.,
  Hale S.~J.,  Miller B.,   New R.,  2009, \mn@doi [MNRAS: Letters]
  {10.1111/j.1745-3933.2009.00672.x}, 396, L100

\bibitem[\protect\citeauthoryear{{Brown}}{{Brown}}{1984}]{Brown1984}
{Brown} T.~M.,  1984, \mn@doi [Science] {10.1126/science.226.4675.687}, 226,
  687

\bibitem[\protect\citeauthoryear{Cassisi, Salaris  \& Irwin}{Cassisi
  et~al.}{2003}]{Cassisi2003}
Cassisi S.,  Salaris M.,   Irwin A.~W.,  2003, \mn@doi [ApJ] {10.1086/374218},
  588, 862

\bibitem[\protect\citeauthoryear{Christensen-Dalsgaard}{Christensen-Dalsgaard}{2008}]{Christensen-Dalsgaard2008a}
Christensen-Dalsgaard J.,  2008, \mn@doi [Ap\&SS] {10.1007/s10509-007-9689-z},
  316, 113

\bibitem[\protect\citeauthoryear{Christensen-Dalsgaard, D{\"{a}}ppen  \&
  Lebreton}{Christensen-Dalsgaard et~al.}{1988}]{Christensen-Dalsgaard1988}
Christensen-Dalsgaard J.,  D{\"{a}}ppen W.,   Lebreton Y.,  1988, \mn@doi
  [Nature] {10.1038/336634a0}, 336, 634

\bibitem[\protect\citeauthoryear{Davies, Broomhall, Chaplin, Elsworth  \&
  Hale}{Davies et~al.}{2014}]{Davies2014}
Davies G.~R.,  Broomhall A.-M.,  Chaplin W.~J.,  Elsworth Y.,   Hale S.~J.,
  2014, \mn@doi [MNRAS] {10.1093/mnras/stu080}, 439, 2025

\bibitem[\protect\citeauthoryear{{Ferguson}, {Alexander}, {Allard}, {Barman},
  {Bodnarik}, {Hauschildt}, {Heffner-Wong}  \& {Tamanai}}{{Ferguson}
  et~al.}{2005}]{Ferguson2005}
{Ferguson} J.~W.,  {Alexander} D.~R.,  {Allard} F.,  {Barman} T.,  {Bodnarik}
  J.~G.,  {Hauschildt} P.~H.,  {Heffner-Wong} A.,   {Tamanai} A.,  2005,
  \mn@doi [ApJ] {10.1086/428642}, 623, 585

\bibitem[\protect\citeauthoryear{Houdek, Trampedach, Aarslev  \&
  Christensen-Dalsgaard}{Houdek et~al.}{2017}]{Houdek2017}
Houdek G.,  Trampedach R.,  Aarslev M.~J.,   Christensen-Dalsgaard J.,  2017,
  \mn@doi [MNRAS: Letters] {10.1093/mnrasl/slw193}, 464, L124

\bibitem[\protect\citeauthoryear{Iglesias \& Rogers}{Iglesias \&
  Rogers}{1996}]{Iglesias1996}
Iglesias C.~A.,  Rogers F.~J.,  1996, \mn@doi [ApJ] {10.1086/177381}, 464, 943

\bibitem[\protect\citeauthoryear{J{\o}rgensen, Weiss, Mosumgaard, {Silva
  Aguirre}  \& Sahlholdt}{J{\o}rgensen et~al.}{2017}]{Joergensen2017}
J{\o}rgensen A. C.~S.,  Weiss A.,  Mosumgaard J.~R.,  {Silva Aguirre} V.,
  Sahlholdt C.~L.,  2017, \mn@doi [MNRAS] {10.1093/mnras/stx2226}, 472, 3264

\bibitem[\protect\citeauthoryear{Kjeldsen, Bedding  \&
  Christensen-Dalsgaard}{Kjeldsen et~al.}{2008}]{Kjeldsen2008}
Kjeldsen H.,  Bedding T.~R.,   Christensen-Dalsgaard J.,  2008, \mn@doi [ApJ]
  {10.1086/591667}, 683, L175

\bibitem[\protect\citeauthoryear{Magic \& Weiss}{Magic \&
  Weiss}{2016}]{Magic2016}
Magic Z.,  Weiss A.,  2016, \mn@doi [A\&A] {10.1051/0004-6361/201527690}, 592,
  A24

\bibitem[\protect\citeauthoryear{Magic, Collet, Asplund, Trampedach, Hayek,
  Chiavassa, Stein  \& Nordlund}{Magic et~al.}{2013}]{Magic2013}
Magic Z.,  Collet R.,  Asplund M.,  Trampedach R.,  Hayek W.,  Chiavassa A.,
  Stein R.~F.,   Nordlund {\AA}.,  2013, \mn@doi [A\&A]
  {10.1051/0004-6361/201321274}, 557, A26

\bibitem[\protect\citeauthoryear{{Mosumgaard}, {Ball}, {Silva Aguirre}, {Weiss}
   \& {Christensen-Dalsgaard}}{{Mosumgaard} et~al.}{2018}]{Mosumgaard2018a}
{Mosumgaard} J.~R.,  {Ball} W.~H.,  {Silva Aguirre} V.,  {Weiss} A.,
  {Christensen-Dalsgaard} J.,  2018, \mn@doi [MNRAS] {10.1093/mnras/sty1442},
  478, 5663

\bibitem[\protect\citeauthoryear{Piau, Collet, Stein, Trampedach, Morel  \&
  Turck-Chi{\`{e}}ze}{Piau et~al.}{2014}]{Piau2014}
Piau L.,  Collet R.,  Stein R.~F.,  Trampedach R.,  Morel P.,
  Turck-Chi{\`{e}}ze S.,  2014, \mn@doi [MNRAS] {10.1093/mnras/stt1866}, 437,
  164

\bibitem[\protect\citeauthoryear{Rosenthal, Christensen-Dalsgaard, Nordlund,
  Stein  \& Trampedach}{Rosenthal et~al.}{1999}]{Rosenthal1999}
Rosenthal C.~S.,  Christensen-Dalsgaard J.,  Nordlund {\AA}.,  Stein R.~F.,
  Trampedach R.,  1999, Astronomy and Astrophysics, 351, 689

\bibitem[\protect\citeauthoryear{Schlattl, Weiss  \& Ludwig}{Schlattl
  et~al.}{1997}]{Schlattl1997}
Schlattl H.,  Weiss A.,   Ludwig H.-G.,  1997, Astronomy and Astrophysics, 322,
  646

\bibitem[\protect\citeauthoryear{{Serenelli}, {Basu}, {Ferguson}  \&
  {Asplund}}{{Serenelli} et~al.}{2009}]{Serenelli2009}
{Serenelli} A.~M.,  {Basu} S.,  {Ferguson} J.~W.,   {Asplund} M.,  2009,
  \mn@doi [ApJ: Letters] {10.1088/0004-637X/705/2/L123}, 705, L123

\bibitem[\protect\citeauthoryear{Sonoi, Samadi, Belkacem, Ludwig, Caffau  \&
  Mosser}{Sonoi et~al.}{2015}]{Sonoi2015}
Sonoi T.,  Samadi R.,  Belkacem K.,  Ludwig H.-G.,  Caffau E.,   Mosser B.,
  2015, \mn@doi [A\&A] {10.1051/0004-6361/201526838}, 583, A112

\bibitem[\protect\citeauthoryear{Trampedach, Stein, Christensen-Dalsgaard,
  Nordlund  \& Asplund}{Trampedach et~al.}{2014a}]{Trampedach2014a}
Trampedach R.,  Stein R.~F.,  Christensen-Dalsgaard J.,  Nordlund {\AA}.,
  Asplund M.,  2014a, \mn@doi [MNRAS] {10.1093/mnras/stu889}, 442, 805

\bibitem[\protect\citeauthoryear{Trampedach, Stein, Christensen-Dalsgaard,
  Nordlund  \& Asplund}{Trampedach et~al.}{2014b}]{Trampedach2014b}
Trampedach R.,  Stein R.~F.,  Christensen-Dalsgaard J.,  Nordlund {\AA}.,
  Asplund M.,  2014b, \mn@doi [MNRAS] {10.1093/mnras/stu2084}, 445, 4366

\bibitem[\protect\citeauthoryear{Trampedach, Aarslev, Houdek, Collet,
  Christensen-Dalsgaard, Stein  \& Asplund}{Trampedach
  et~al.}{2017}]{Trampedach2017}
Trampedach R.,  Aarslev M.~J.,  Houdek G.,  Collet R.,  Christensen-Dalsgaard
  J.,  Stein R.~F.,   Asplund M.,  2017, \mn@doi [MNRAS]
  {10.1093/mnrasl/slw230}, 466, L43

\bibitem[\protect\citeauthoryear{Weiss \& Schlattl}{Weiss \&
  Schlattl}{2008}]{Weiss2008}
Weiss A.,  Schlattl H.,  2008, \mn@doi [Ap\&SS] {10.1007/s10509-007-9606-5},
  316, 99

\makeatother
\end{thebibliography}

\bsp	
\label{lastpage}
\end{document}